\documentclass[preprint2]{aastex}
\usepackage{spr-astr-addons}
\usepackage{url}
\bibpunct{(}{)}{,}{n}{,}{,}

\RequirePackage{color}

\begin{document}

 \title{A trio of horseshoes: past, present and future dynamical 
        evolution of Earth co-orbital asteroids 2015~XX$_{169}$, 
        2015~YA and 2015~YQ$_{1}$ 
        }

 \shorttitle{Dynamics of 2015~XX$_{169}$, 2015~YA and 2015~YQ$_{1}$}
 \shortauthors{de la Fuente Marcos and de la Fuente Marcos}

 \author{C.~de~la~Fuente Marcos} 
  \and 
 \author{R.~de~la~Fuente Marcos} 
 \affil{Apartado de Correos 3413, E-28080 Madrid, Spain} 
 \email{carlosdlfmarcos@gmail.com}

 \begin{abstract}
    It is widely accepted that a quasi-steady-state flux of 
    minor bodies moving in and out of the co-orbital state 
    with the Earth may exist. Some of these objects are very 
    good candidates for future in situ study due to their 
    favourable dynamical properties. In this paper, we show 
    that the recently discovered near-Earth asteroids 
    2015~XX$_{169}$, 2015~YA and 2015~YQ$_{1}$ are small 
    transient Earth co-orbitals. These new findings increase 
    the tally of known Earth co-orbitals to 17. The three of 
    them currently exhibit asymmetric horseshoe behaviour 
    subjected to a Kozai resonance and their short-term 
    orbital evolution is rather unstable. Both 2015~YA and 
    2015~YQ$_{1}$ may leave Earth's co-orbital zone in the 
    near future as they experience close encounters with 
    Venus, the Earth-Moon system and Mars. Asteroid 
    2015~XX$_{169}$ may have remained in the vicinity of, or 
    trapped inside, the 1:1 mean motion resonance with our 
    planet for many thousands of years and may continue in 
    that region for a significant amount of time into the 
    future.
 \end{abstract}

 \keywords{Celestial mechanics $\cdot$ 
           Minor planets, asteroids: general $\cdot$
           Minor planets, asteroids: individual: 2015~XX$_{169}$ $\cdot$
           Minor planets, asteroids: individual: 2015~YA $\cdot$
           Minor planets, asteroids: individual: 2015~YQ$_{1}$ $\cdot$
           Planets and satellites: individual: Earth 
          }

 \section{Introduction}
    During the last two decades, observations of near-Earth Objects (NEOs) have uncovered the existence of a transient near-Earth asteroid 
    (NEA) population that goes around the Sun in almost exactly one Earth's orbital period, i.e., they are trapped in the 1:1 mean motion 
    resonance with our planet. Although their orbits may not be at all similar to that of the Earth if they are very eccentric and inclined, 
    these peculiar NEAs are still referred to in the literature as Earth co-orbitals (Morais \& Morbidelli 2002). Some of these NEAs may 
    have had their origin in the Earth-Moon system (see, e.g., Margot \& Nicholson 2003) although other sources in the main asteroid belt 
    are perhaps more likely (see, e.g., Morais \& Morbidelli 2002). However, the effects derived from orbital chaos severely limit our 
    ability to determine their sources in a reliable manner (Connors et al. 2004). 

    It is widely accepted that a quasi-steady-state flux of minor bodies moving in and out of the co-orbital state with the Earth may exist 
    (see, e.g., Morais \& Morbidelli 2002). The list of documented Earth co-orbitals now includes 14 objects (see, e.g., de la Fuente Marcos 
    \& de la Fuente Marcos 2016). Most of these minor bodies have diameters smaller than 50~m (which is equivalent to an absolute magnitude, 
    $H$, in the range 23.5--25.5 for an assumed albedo in the range 0.20--0.04). Some of these NEOs are relatively easy to access from our 
    planet, experiencing close flybys (i.e., have short perigee distances), and as such they have been included in the Near-Earth Object 
    Human Space Flight Accessible Targets Study (NHATS)\footnote{\url{http://neo.jpl.nasa.gov/nhats/}} list (Abell et~al. 2012a,b). The 
    actual size of this interesting population could be large because small bodies probably dominate this dynamical class.  

    Unfortunately and for NEOs, being small is usually associated with not getting much attention. A large number of NEO candidates listed 
    on the Minor Planet Center's (MPC) NEO Confirmation Page\footnote{\url{http://www.minorplanetcenter.net/iau/NEO/toconfirm_tabular.html}} 
    attract no follow up if their preliminary orbital data indicate that they are probably very small. Follow-up observers select targets
    that their telescope can observe, and often small objects have a very limited window of observability, thus do not get observed as much
    as larger objects. It is frequently the case that such small objects do not accumulate enough observations even to be formally 
    considered discovered and consequently they do not get a designation at all. Therefore, they become lost asteroids at discovery time. 
    Even for those that have been officially discovered (i.e., received a designation), being small almost certainly implies that their 
    orbits will remain poorly determined for decades. For instance, the number of recovered objects with $H>25.5$~mag in the extensive study 
    performed by Harris \& D'Abramo (2015) is zero and only for objects with $H<20$~mag the recovery probability (that of observing the 
    object again some period, often a year, after discovery and, consequently, improve its orbital solution) was found to be greater than 
    50\%. On the other hand, recovering a small NEO not only depends on its apparent magnitude at perigee ---that must be $<22$ in $V$ for 
    most active surveys--- but on its rate of motion as well because of image trailing loss of very small, fast moving bodies. For example, 
    a NEO travelling faster than 10\degr per day must reach an apparent magnitude at perigee $<20$ in order to be detected (see fig. 1 in 
    Harris \& D'Abramo 2015). The faintest asteroid observation performed so far corresponded to $V=26.7$~mag for asteroid 2008 LG$_{2}$ 
    (Micheli et al. 2015), but it was accomplished by using targeted follow-up with the telescope tracking rate matching closely the 
    asteroid's angular motion in the sky. Currently active and past NEO search programs do not have the ability to go as faint as 
    $V\sim25$~mag in survey mode (Farnocchia et al. 2016). This means in practice that most recoveries of small NEOs are serendipitous 
    rather than planned.  

    Given the observational challenges pointed out above, many small Earth co-orbitals may get lost soon after being discovered and the ones 
    receiving an official designation are strong candidates to join the group of asteroids with poorly determined orbits waiting for an 
    uncertain recovery. This state of affairs has a negative impact on our understanding of how a quasi-steady-state flux of minor bodies 
    moving in and out of the co-orbital state with the Earth may operate. Here, we show that the recently discovered small NEAs 
    2015~XX$_{169}$, 2015~YA and 2015~YQ$_{1}$ are performing the corkscrew motion characteristic of Earth co-orbitals of the horseshoe 
    type. With these three discoveries the number of known Earth co-orbitals increases to 17 with 12 of them following some kind of 
    horseshoe path with respect to our planet. Many of them are subjected to a Kozai resonance (Kozai 1962). This paper is organized as 
    follows. In Sect. 2, we discuss the methodology followed in this work. In Sect. 3, we present the available data on 2015~XX$_{169}$ and 
    study its dynamical evolution as well as the impact of errors on our results. Sections 4 and 5 are equivalent to Sect. 3 but for 2015~YA 
    and 2015~YQ$_{1}$, respectively. In Sect. 6, we discuss the comparative dynamics of these objects and place them within the context of 
    the previously known Earth co-orbitals. Our conclusions are presented in Sect. 7.

 \section{Earth co-orbitals: what are they and how to study them}
    Members of the population of NEAs in co-orbital motion with our planet are not characterised as such at discovery time but identified 
    a posteriori during the analysis of their past, present and future orbital evolution that is explored using $N$-body simulations. 

    \subsection{Characterising co-orbital objects}
       Co-orbital objects (asteroids or comets) move inside the 1:1 mean motion resonance with a planet. As such, they go around a certain 
       star ---e.g., the Sun in the case of the Earth--- in almost exactly one orbital period of their host planet. In contrast with natural 
       satellites, they do not follow closed planetocentric paths but from the planet's point of view they loop around, in some cases for 
       billions of years. 

       Co-orbital bodies are identified numerically by studying the behaviour of their relative mean longitude, $\lambda_{\rm r}$, or 
       difference between the mean longitude of the object and that of its host planet. The relative mean longitude of a passing body with 
       respect to a given planet can take any value in the interval (0, 360)\degr, i.e. circulates; in sharp contrast, the relative mean 
       longitude of co-orbital bodies librates or oscillates around a certain value (in principle, 0\degr, $\pm$60\degr or 180\degr, but the 
       actual value depends on the orbital eccentricity and inclination). The mean longitude of an object ---planet, asteroid or comet--- is 
       given by $\lambda=M+\Omega+\omega$, where $M$ is the mean anomaly, $\Omega$ is the longitude of the ascending node, and $\omega$ is 
       the argument of perihelion (see, e.g., Murray \& Dermott 1999). 

       The degree of complexity of the co-orbital dynamics experienced by an object in a multi-planet environment could be very high. In 
       addition to the three elementary configurations ---quasi-satellite or retrograde satellite, Trojan or tadpole, and horseshoe--- 
       compound states (see, e.g., Morais \& Morbidelli 2002) as well as recurrent transitions between the various configurations (Namouni 
       et al. 1999; Namouni \& Murray 2000) are possible. In the Solar System and focusing on long-term stability, the most likely 
       configuration is by far the Trojan state that ---for low values of both eccentricity and inclination--- is characterised by the 
       libration of $\lambda_{\rm r}$ around $\pm60\degr$ as the affected minor body follows a tadpole orbit with respect to the host 
       planet (see, e.g., Murray \& Dermott 1999); such an object is classified as an $L_4$ Trojan when the value of $\lambda_{\rm r}$ 
       oscillates around +60\degr, or as an $L_5$ Trojan if the value librates around $-60\degr$ (or 300\degr). For eccentric orbits, the 
       values of $\lambda_{\rm r}$ are displaced towards 180\degr by an amount proportional to the orbital eccentricity (Namouni 1999; 
       Namouni et al. 1999; Namouni \& Murray 2000; Nesvorny et al. 2002; Morais \& Morbidelli 2002). If instead of stability we are more 
       concerned about plain likelihood, then the most probable co-orbital state is the one characterised by a libration amplitude larger 
       than 180\degr around a value of $\lambda_{\rm r}=180\degr$ and often enclosing $\pm60\degr$; such objects follow horseshoe orbits 
       (see, e.g., Murray \& Dermott 1999). At regular intervals and before moving away, a horseshoe librator follows a corkscrew-like 
       trajectory in the vicinity of its host planet for a certain period of time, a decade or more in the case of the Earth (see, e.g., de 
       la Fuente Marcos \& de la Fuente Marcos 2016). Far less likely than the horseshoe configuration, the quasi-satellite dynamical state 
       was first described by Jackson (1913). In this case, the object involved appears to travel around the planet but is not 
       gravitationally bound to it: the body librates around the longitude of its associated planet ---i.e., $\lambda_{\rm r}$ oscillates 
       around 0\degr--- but its trajectory is not closed (see, e.g., Mikkola et al. 2006).
%
%
     \begin{table*}
       \centering
        \fontsize{8}{11pt}\selectfont
        \tabcolsep 0.35truecm
        \caption{Heliocentric ecliptic Keplerian orbital elements of NEAs 2015~XX$_{169}$, 2015~YA and 2015~YQ$_{1}$. Values include the 
                 1$\sigma$ uncertainty (Epoch = JD2457400.5, 2016-January-13.0; J2000.0 ecliptic and equinox. Source: JPL Small-Body 
                 Database.)                 
                }
        \begin{tabular}{lllll}
         \hline
                                                            &   &       2015~XX$_{169}$ &               2015~YA &         2015~YQ$_{1}$ \\
         \hline
          Semi-major axis, $a$ (AU)                         & = &   0.99974$\pm$0.00003 &   0.99753$\pm$0.00005 &   1.00134$\pm$0.00005 \\
          Eccentricity, $e$                                 & = &   0.18364$\pm$0.00012 &   0.2791$\pm$0.0002   &   0.40398$\pm$0.00013 \\
          Inclination, $i$ (\degr)                          & = &   7.691$\pm$0.006     &   1.6249$\pm$0.0009   &   2.4865$\pm$0.0010   \\
          Longitude of the ascending node, $\Omega$ (\degr) & = & 256.7497$\pm$0.0002   & 255.3291$\pm$0.0004   &  88.89770$\pm$0.00004 \\
          Argument of perihelion, $\omega$ (\degr)          & = & 283.918$\pm$0.002     &  83.849$\pm$0.002     & 112.185$\pm$0.002     \\
          Mean anomaly, $M$ (\degr)                         & = & 310.655$\pm$0.012     &  99.79$\pm$0.02       & 317.067$\pm$0.012     \\
          Perihelion, $q$ (AU)                              & = &   0.81614$\pm$0.00010 &   0.71908$\pm$0.00012 &   0.59681$\pm$0.00010 \\
          Aphelion, $Q$ (AU)                                & = &   1.18333$\pm$0.00004 &   1.27598$\pm$0.00006 &   1.40586$\pm$0.00008 \\
          Absolute magnitude, $H$ (mag)                     & = &  27.4$\pm$0.4         &  27.4$\pm$0.3         &  28.1$\pm$0.5         \\
         \hline
        \end{tabular}
        \label{elements}
     \end{table*}
%
%

    \subsection{Numerical investigation}
       As pointed out above, co-orbital objects are identified indirectly after investigating numerically the evolution of their relative 
       mean longitude with respect to a given planet. In this work and following de la Fuente Marcos \& de la Fuente Marcos (2012), we 
       perform extensive direct $N$-body calculations using the Hermite scheme described by Makino (1991) and implemented by Aarseth (2003). 
       The standard version of this direct $N$-body code is publicly available from the IoA web 
       site.\footnote{\url{http://www.ast.cam.ac.uk/~sverre/web/pages/nbody.htm}} Initial conditions (positions and velocities in the 
       barycentre of the Solar System) have been obtained from the Jet Propulsion Laboratory (JPL) HORIZONS system (Giorgini et al. 1996; 
       Standish 1998) and they are referred to the epoch JD 2457400.5 (2016-January-13.0), which is the $t$ = 0 instant in our figures. Our 
       calculations take into account perturbations from the eight major planets, the Moon, the barycentre of the Pluto-Charon system, and 
       the three largest asteroids. Non-gravitational forces, relativistic and oblateness terms were not included in the calculations.

       The current orbital solutions of the three small NEAs ---2015~XX$_{169}$, 2015~YA and 2015~YQ$_{1}$--- studied here are relatively 
       poor (see Table \ref{elements}) as they are based on short arcs but their degrees of uncertainty are similar to that of the recently 
       identified Earth co-orbital 2015~SO$_{2}$ (de la Fuente Marcos \& de la Fuente Marcos 2016). In such cases, a statistical approach 
       can be used to evaluate the current dynamical state of the minor body probabilistically. In order to do that and in addition to 
       making use of the nominal orbit in Table \ref{elements}, we have computed 50 control simulations for $\pm$50 kyr for each object with 
       sets of orbital elements obtained from the nominal ones within the quoted uncertainties and assuming Gaussian distributions for them 
       up to $\pm9\sigma$. For instance, a new value of the semi-major axis for a control orbit has been found using the expression 
       $a_{\rm t}=\langle{a}\rangle+n\ \sigma_{a}\,r_{\rm i}$, where $a_{\rm t}$ is the semi-major axis of the test orbit, 
       $\langle{a}\rangle$ is the mean value of the semi-major axis from the available orbit (Table \ref{elements}), $n$ is a suitable 
       integer (in our case, 9), $\sigma_{a}$ is the standard deviation of $a$ (Table \ref{elements}), and $r_{\rm i}$ is a (pseudo) random 
       number with normal distribution in the range $-$1 to 1. In the figures when an orbit is labelled `$\pm{n}\sigma$', where $n$ is an 
       integer, it has been obtained by adding (+) or subtracting ($-$) $n$-times the uncertainty from the orbital parameters (the six 
       elements) in Table \ref{elements}. In addition, we have computed two sets of 100 shorter control simulations in both directions of 
       time and analysed the short-term evolution of the orbital elements $a$, $e$, $i$, $\Omega$ and $\omega$ for each object studied here 
       (within 1$\sigma$ and 9$\sigma$, respectively) focusing on the average values and the ranges (minimum and maximum) of the parameters.

       Sitarski (1998, 1999, 2006) has pointed out that the procedure used above is equivalent to considering a number of different virtual
       objects moving in similar orbits, but not a sample of test orbits derived from a set of observations obtained for a single object. In 
       other words, given a set of observations of a certain object, the values of the computed orbital elements are not mutually 
       independent. The correct statistical alternative is to consider how the elements affect each other, applying the Monte Carlo using 
       the Covariance Matrix (MCCM) approach (Bordovitsyna et al. 2001; Avdyushev \& Banschikova 2007), or to follow the procedure described 
       in Sitarski (1998, 1999, 2006). As a consistency test, we have used the implementation of the MCCM approach described in de la Fuente 
       Marcos \& de la Fuente Marcos (2015b) to recompute the orbital evolution of the objects studied producing an additional set of 100 
       short control simulations similar to the other two pointed out above. The respective covariance matrices have been obtained from the 
       JPL Small-Body Database.

       The study of the past, present and future evolution of hundreds of control orbits for each object using both classical and MCCM 
       techniques confirms the robustness of the objects' current resonant state as derived from our calculations and the relatively poor 
       orbits available. Based on the number of computations performed, we can state that 2015~XX$_{169}$, 2015~YA and 2015~YQ$_{1}$ are 
       current co-orbitals of the Earth with a probability $>99.8$\%. However, all of them exhibit a rather unstable orbital evolution and 
       two of them may not remain within Earth's co-orbital region for long. The time-scale for exponential divergence of initially close 
       orbits or Lyapunov time of these objects is in some cases as short as a few decades.

%
%
     \begin{figure}
       \centering
        \includegraphics[width=\linewidth]{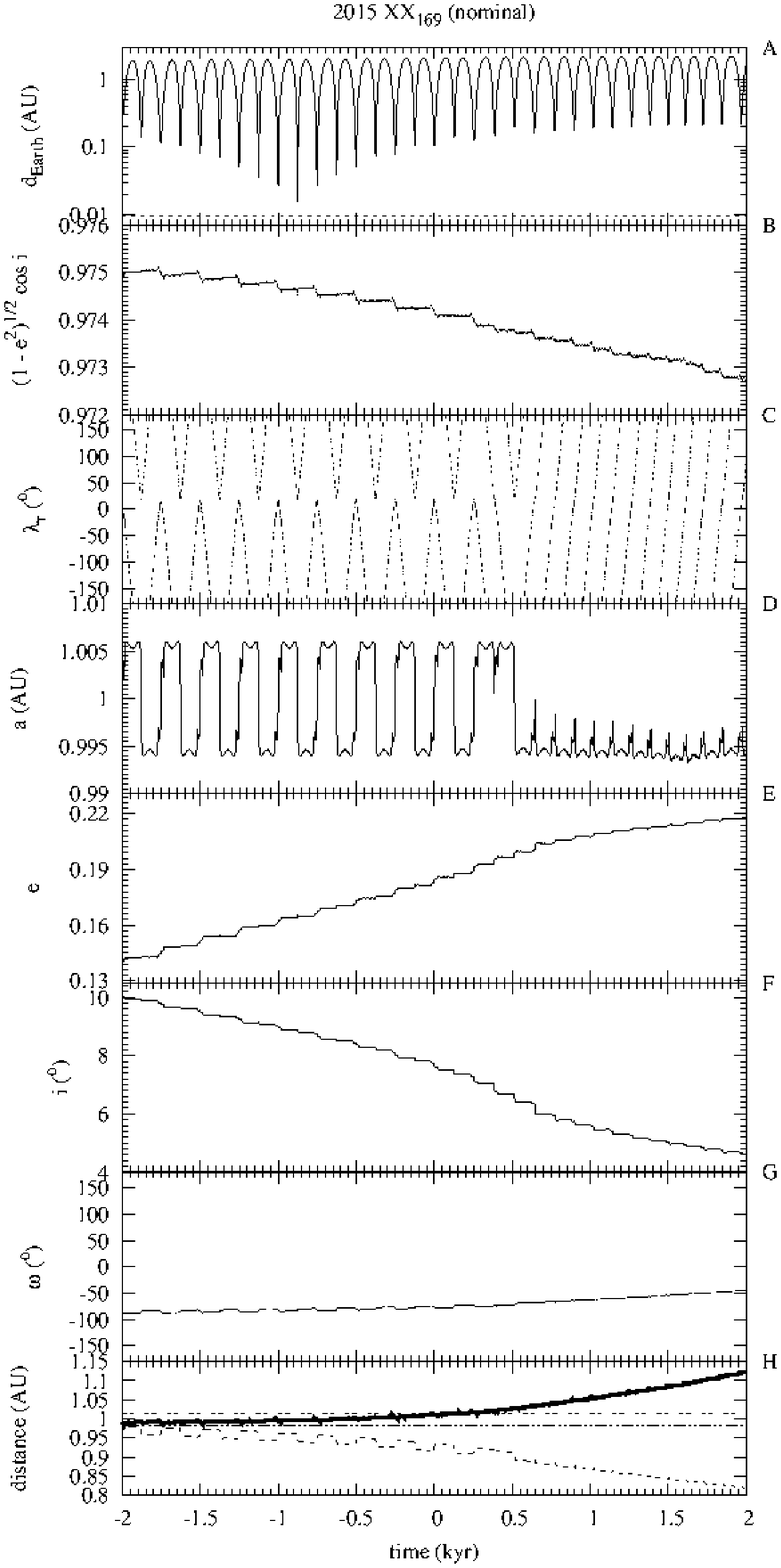}
        \caption{Time evolution of various parameters for the nominal orbit of 2015~XX$_{169}$ during the time interval (-2000, 2000) yr. 
                 The distance from the Earth (panel A) with the value of the radius of the Hill sphere of the Earth, 0.0098 AU (dashed line). 
                 The parameter $\sqrt{1 - e^2} \cos i$ (panel B). The resonant angle, $\lambda_{\rm r}$ (panel C). The orbital elements $a$ 
                 (panel D), $e$ (panel E), $i$ (panel F), and $\omega$ (panel G). The distances to the descending (thick line) and ascending 
                 nodes (dotted line) are plotted in panel H; Earth's aphelion and perihelion distances are also shown.
                }
        \label{shortxx169}
     \end{figure}
%
%
 \section{Asteroid 2015~XX$_{169}$: data and results}
    Asteroid 2015~XX$_{169}$ was discovered on 2015 December 9 by R. G. Matheny observing for the Mt. Lemmon Survey with the 1.5-m reflector
    of the programme (Stecklum et al. 2015). It was first observed at $V=20.9$~mag and the collected data showed that it is a very small 
    object with an absolute magnitude, $H$, of 27.4; if a value of the albedo in the range 0.20--0.04 is assumed, this absolute magnitude is 
    equivalent to a diameter in the range 9--22 m. It has an Earth Minimum Orbit Intersection Distance (MOID) of 0.015 AU and it has been 
    included in the NHATS list. Its orbit is in need of further observations as it is currently based on 37 data points acquired during 4 d 
    (see Table \ref{elements}). Although this object is a typical example of the problematic situation described above, small object with 
    relatively few observations, we will show that in this case it is still possible to arrive to statistically robust conclusions regarding 
    its current dynamical state using the available orbit. 

    At present, 2015~XX$_{169}$ belongs to the Aten dynamical class and moves in an orbit with a value of the semi-major axis of 0.99974 AU,
    similar to that of our planet (1.00074 AU), eccentricity of 0.18364, and moderate inclination, $i$ = 7\fdg7. The values of the 
    Heliocentric Keplerian osculating orbital elements and their uncertainties in Table \ref{elements} (as provided by the JPL Small-Body 
    Database\footnote{\url{http://ssd.jpl.nasa.gov/sbdb.cgi}}) are referred to the epoch JD2457400.5, i.e. a time after its discovery and 
    subsequent close encounter with our planet on 2015 December 14. As a result of this event, the value of the semi-major axis of 
    2015~XX$_{169}$ is slowly drifting upwards and the asteroid will become an Apollo asteroid within less than a year from now (see Fig. 
    \ref{shortxx169}). After that, the value of the semi-major axis of the asteroid will remain in the range 1.002 AU to 1.006 AU for over 
    130 years to decrease again after the next close encounter with our planet, returning to the Aten dynamical class.

%
%
    \begin{figure*}
      \centering
       \includegraphics[width=0.325\linewidth]{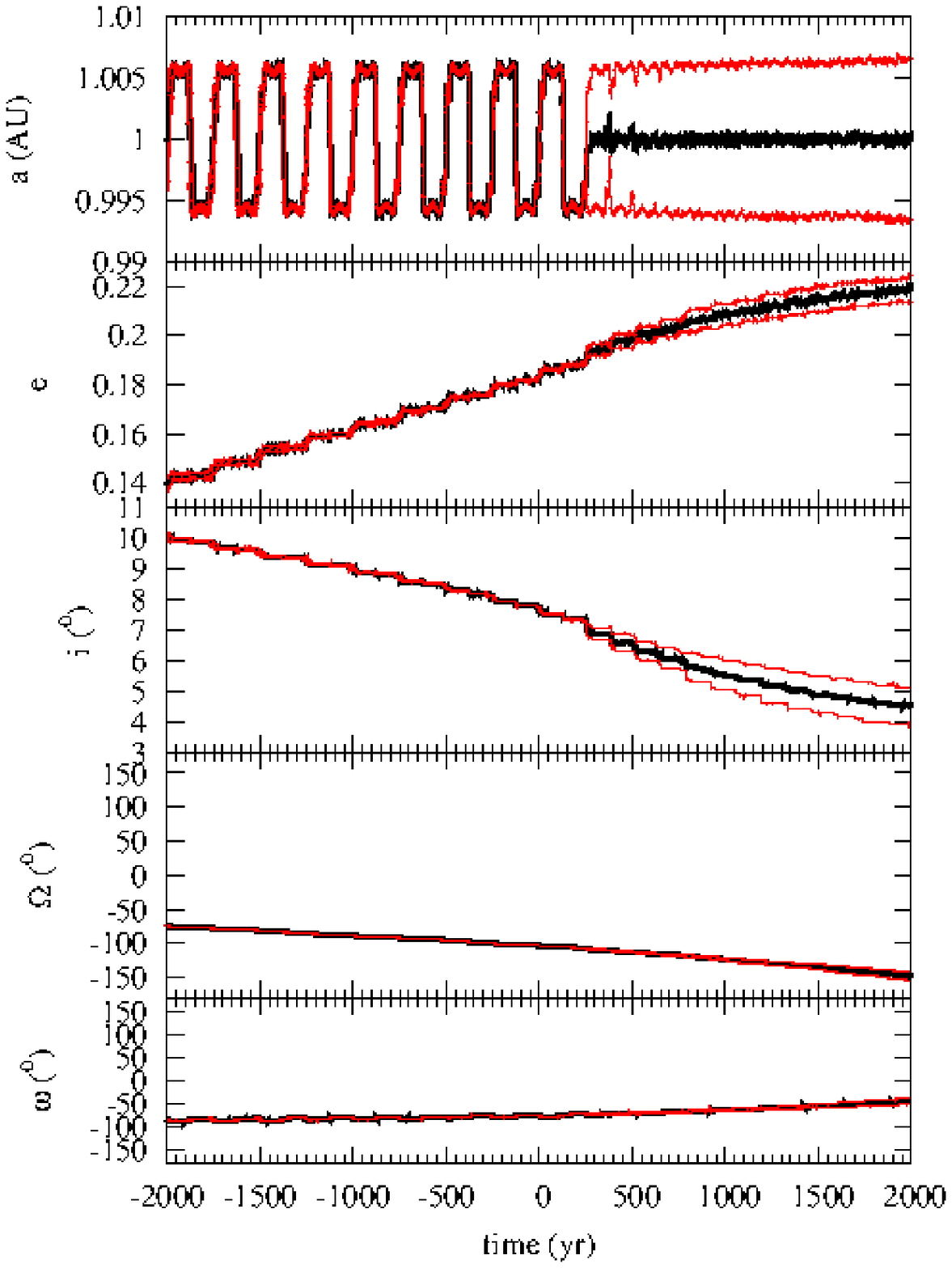}
       \includegraphics[width=0.325\linewidth]{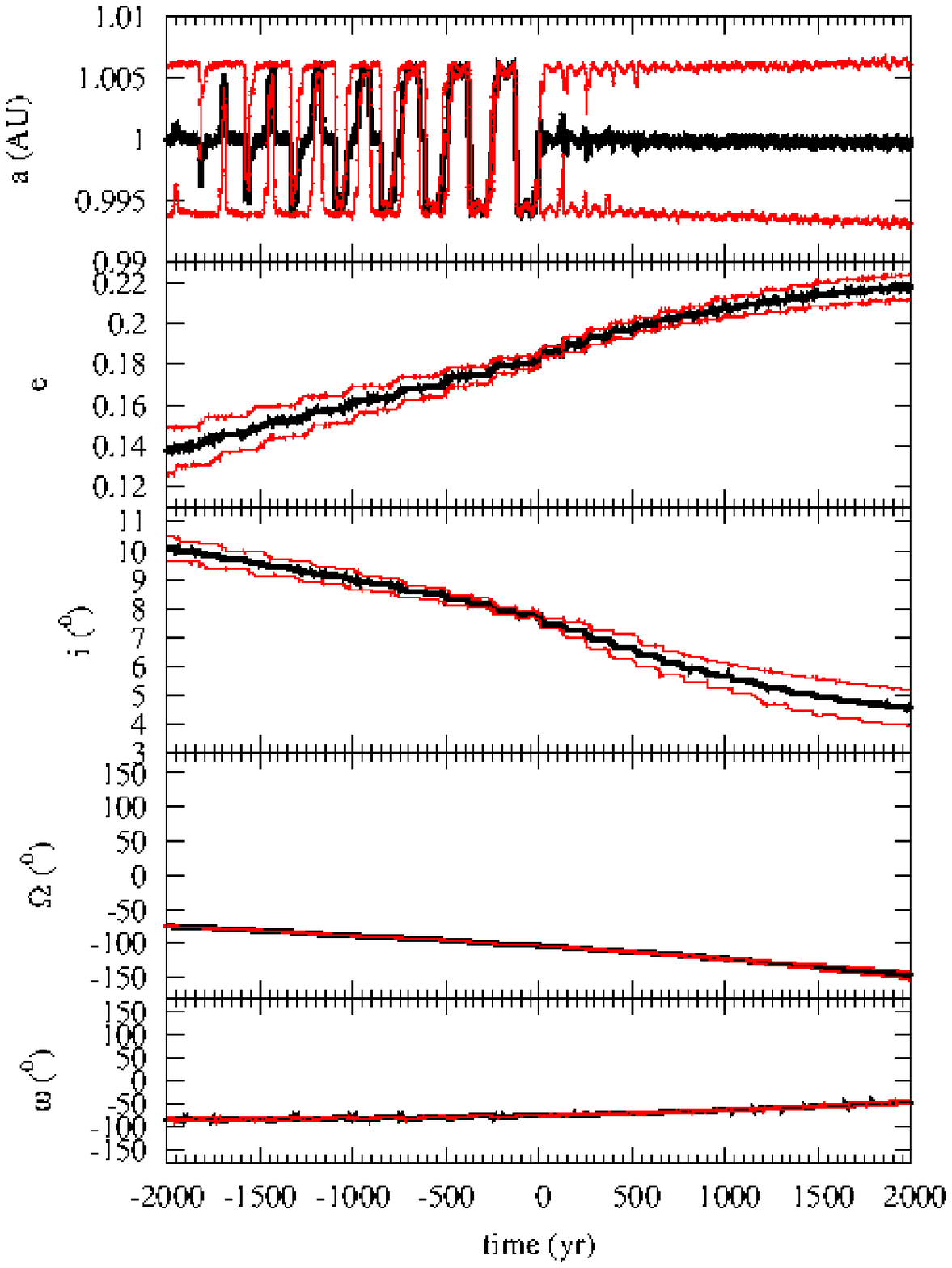}
       \includegraphics[width=0.325\linewidth]{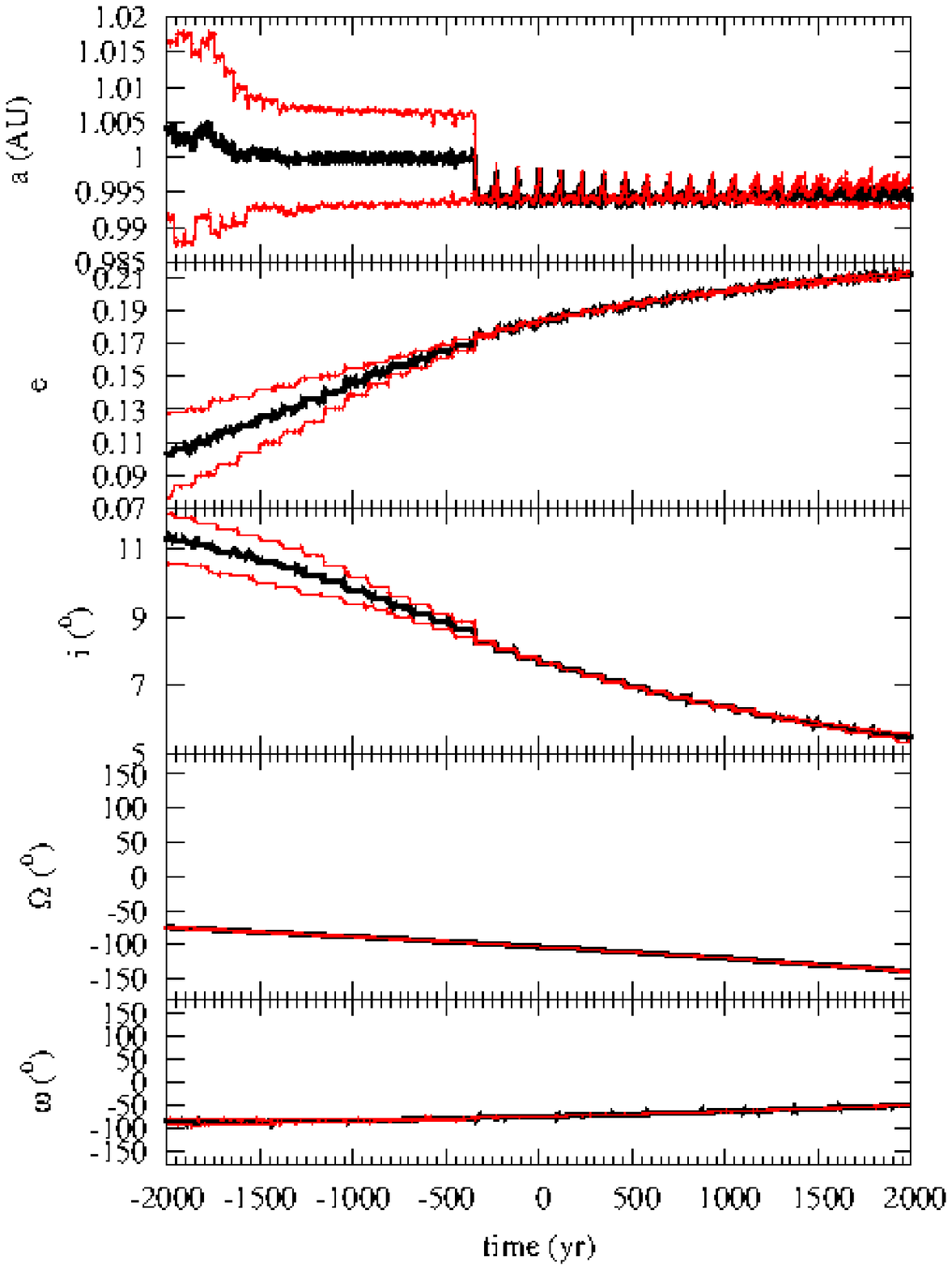}
       \caption{Time evolution of the orbital elements $a$, $e$, $i$, $\Omega$ and $\omega$ of 2015~XX$_{169}$. The black thick curve shows 
                the average evolution of 100 control orbits, the red thin curves show the ranges in the values of the parameters at the 
                given time. Results for a 1$\sigma$ spread in the initial values of the orbital elements (left-hand panels), a 9$\sigma$ 
                spread (central panels), and using MCCM (see the text, right-hand panels).
               }
       \label{errorsxx169}
    \end{figure*}
%
%
    The immediate, past and future, orbital evolution of 2015~XX$_{169}$ is displayed in Fig. \ref{shortxx169}. All the control orbits
    exhibit consistent behaviour within a few hundred years of $t = 0$. Its Lyapunov time was longer in the past, but now it is just a few 
    hundred years. Figure \ref{errorsxx169}, left-hand and central panels, shows that this object was considerably more stable in the past 
    and that its recent close encounter with the Earth has made the orbit significantly less stable although 2015~XX$_{169}$ is not expected 
    to leave the vicinity of Earth's co-orbital zone within tens of thousands of years (see below); based on the integrations performed, 
    the Earth's co-orbital region currently extends from $\sim0.994$~AU to $\sim1.006$~AU. Consistent with the warning conveyed in Sitarski 
    (1998, 1999, 2006), the average, short-term evolution of the path followed by 2015~XX$_{169}$ as derived using the MCCM approach (see 
    Fig. \ref{errorsxx169}, right-hand panels) is rather different ---and more chaotic into the past--- than the one obtained when the 
    classical method is applied. It is clear that the average evolution from the MCCM approach is more in line with that of the nominal 
    orbit plotted in Fig. \ref{shortxx169}. In any case, consistent dynamical evolution is systematically found within reasonable limits (a 
    few hundred years of $t = 0$) for all the control orbits studied. 

    Asteroid 2015~XX$_{169}$ follows an asymmetrical horseshoe trajectory with respect to the Earth; $\lambda_{\rm r}$ librates around 
    180\degr, but enclosing 0\degr (see Fig. \ref{shortxx169}, panel C). The value of $\lambda_{\rm r}$ does not behave as expected of a 
    classical horseshoe librator, where $\pm60\degr$ are enclosed by the trajectory but not 0\degr (see, e.g., Murray \& Dermott 1999). The 
    object switches between the Aten and Apollo dynamical classes at regular intervals of about 100 yrs (see Fig. \ref{shortxx169}, panel D)
    and it will cease these transitions in about 500 yrs from now, remaining as an Aten. The value of the Kozai parameter ---$\sqrt{1 - e^2} 
    \cos i$--- remains approximately constant (see Fig. \ref{shortxx169}, panel B) over the displayed time interval as the value of the 
    eccentricity increases (see Fig. \ref{shortxx169}, panel E) and, conversely, that of the inclination decreases (see Fig. 
    \ref{shortxx169}, panel F). The value of the argument of perihelion remains confined in the range (-100, -50)\degr (see Fig. 
    \ref{shortxx169}, panel G) which is consistent with a Kozai resonance with libration around -90\degr (or 270\degr), although not very 
    strong. At present, the descending node is close to Earth's aphelion and the ascending node is away from the path of any planet (see 
    Fig. \ref{shortxx169}, panel H). The distance between the Sun and the nodes for a prograde orbit is given by
    \begin{equation}
       r=a(1-e^2)/(1\pm{e}\cos\omega)\,, \label{nodeseq}
    \end{equation}
    where the "+" sign is for the ascending node (where the orbit crosses the Ecliptic from South to North) and the "$-$" sign is for the
    descending node. Close encounters are only possible with the Earth and at the descending node; this is the configuration that the object 
    reached on 2015 December 14 during its last approach to our planet.

%
%
     \begin{figure*}
       \centering
        \includegraphics[width=\linewidth]{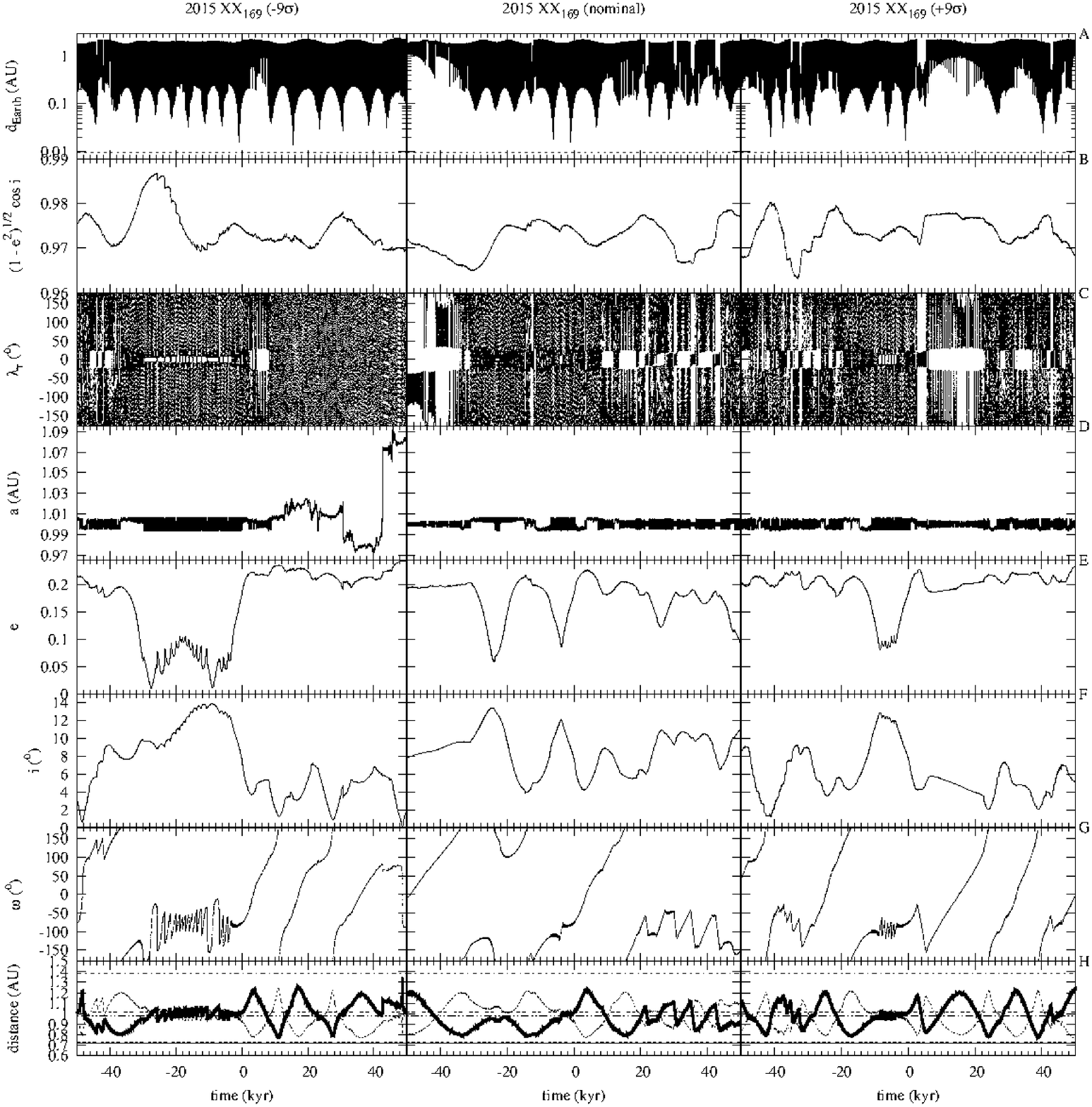}
        \caption{Same as Fig. \ref{shortxx169} but for the nominal orbit (same data as in Fig. \ref{shortxx169} but over a longer time span) 
                 and two representative examples of orbits that are the most different from the nominal one among those integrated up to
                 $\pm9\sigma$ deviations (see the text for details). Venus', Earth's and Mars' perihelion distances, and Venus' and Earth's 
                 aphelion distances are also shown in panel H.
                }
        \label{longxx169}
     \end{figure*}
%
%

    Representative longer integrations are displayed in Fig. \ref{longxx169}. Asteroid 2015~XX$_{169}$ may remain as co-orbital of our 
    planet for many thousands of years into the future and may have remained as such for an equally long period of time, or even longer, in 
    the past. In many calculations (see left-hand and right-hand C panels in Fig. \ref{longxx169}), multiple instances of back and forth 
    transitions between the horseshoe and quasi-satellite resonant states are observed. These transitions are similar to those that 
    characterise the orbital evolution of 2015~SO$_{2}$ as described in de la Fuente Marcos \& de la Fuente Marcos (2016). The mechanism 
    behind these transitions is analogous to the one driving the dynamics of 2015~SO$_{2}$. When this behaviour is observed, the object 
    exhibits Kozai-like dynamics with its nodes confined between Earth's aphelion and perihelion (see panel H in Fig. \ref{longxx169}). In 
    general, any abrupt changes in the state of motion of this object are the result of relatively close encounters with our planet at 
    distances close to one Hill radius (0.0098 AU). During most of the simulated time, both eccentricity and inclination oscillate with the 
    same frequency but out of phase (see Fig. \ref{longxx169}, panels E and F), however the libration is far from regular which suggests 
    that the Kozai resonance is not particularly strong. Consistently, the value of the Kozai parameter (see Fig. \ref{longxx169}, panel B) 
    exhibits irregular oscillations and the argument of perihelion undergoes phases of obvious libration (see Fig. \ref{longxx169}, panel 
    G). Some stages in the orbital evolution of this object are similar to those observed for 2015~SO$_{2}$ as described in de la Fuente
    Marcos \& de la Fuente Marcos (2016). The object exhibits intermitent Kozai-like behaviour throughout the entire time interval with its 
    nodes often confined between Earth's aphelion and perihelion.  
%
%
     \begin{figure}
       \centering
        \includegraphics[width=\linewidth]{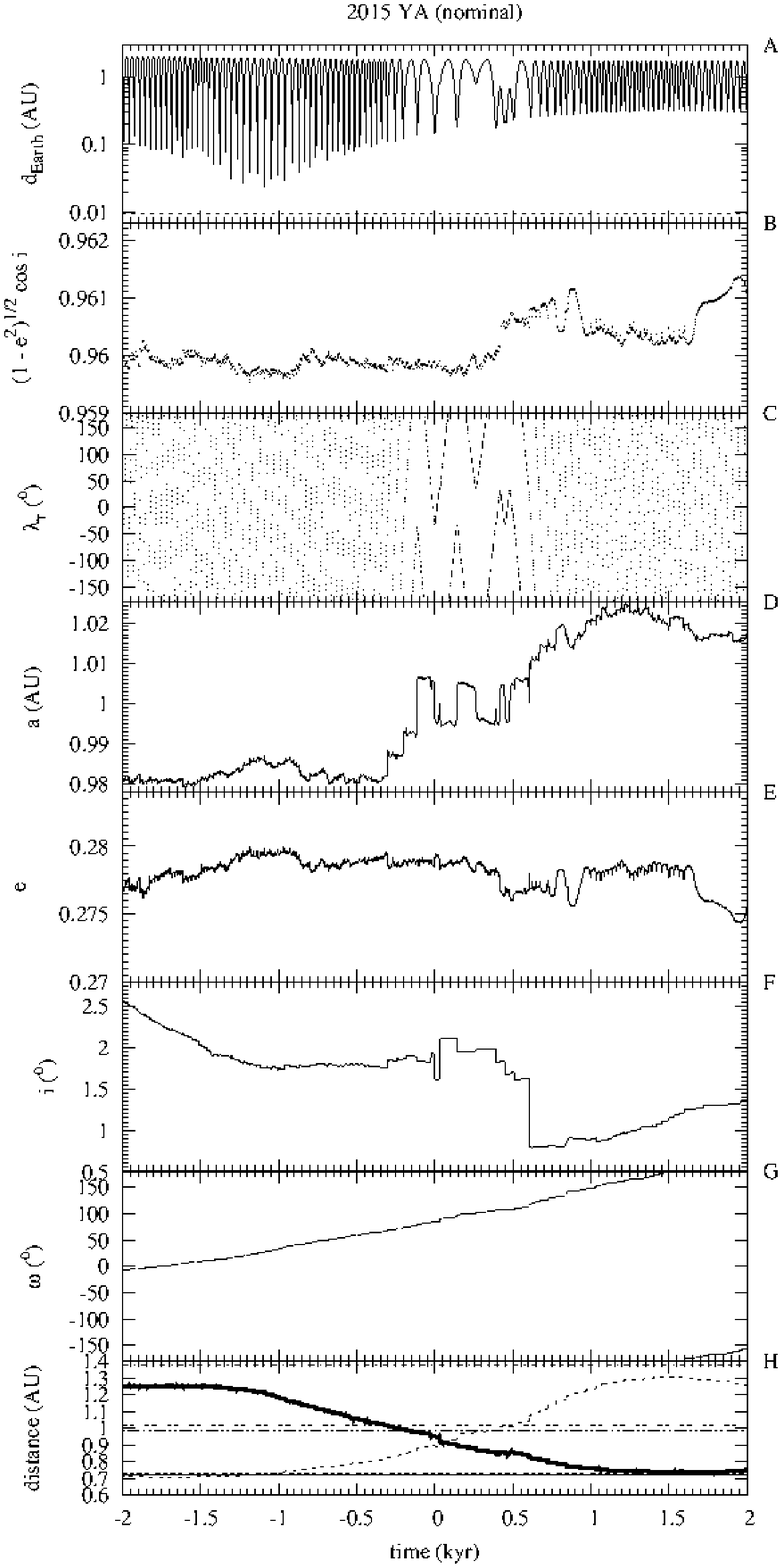}
        \caption{As Fig. \ref{shortxx169} but for 2015~YA. 
                }
        \label{shortya}
     \end{figure}
%
%

 \section{Asteroid 2015~YA: data and results}
    Asteroid 2015~YA was discovered on 2015 December 16 by G. J. Leonard and R. G. Matheny observing for the Catalina Sky Survey (Buzzi et 
    al. 2015) at $V=17.2$~mag. It is as large as 2015~XX$_{169}$ and as such it shares the same problem: small object with relatively few 
    observations. Its current orbital solution is less than satisfactory with 47 observations acquired during 7 d (see Table \ref{elements}), 
    but its uncertainties are similar to those of 2015~XX$_{169}$ or 2015~SO$_{2}$ (de la Fuente Marcos \& de la Fuente Marcos 2016). Its 
    orbital elements show that its path is more eccentric ($e=0.27914$) and less inclined ($i=1\fdg6$) than that of 2015~XX$_{169}$. This 
    opens the possibility of closer encounters with the Earth ---its MOID with our planet has a value of 0.0036 AU--- and even relatively 
    close approaches to both Venus (at perihelion) and Mars (at aphelion). Evolving within this dynamical context, 2015~YA must be less 
    stable than 2015~XX$_{169}$. It is however a member of the NHATS list, NASA's list of viable NEAs for an actual human exploration 
    mission.

    This asteroid is currently a member of the Aten dynamical class but before its recent encounter with the Earth on 2015 December 15 it 
    was an Apollo. It will remain as an Aten for the next 142 years and then return to the Apollo dynamical class. Its short-term dynamical 
    evolution is displayed in Fig. \ref{shortya} (nominal orbit). It arrived very recently to the Earth co-orbital zone and it may leave in 
    a few hundred years. Its orbital evolution is far more chaotic than that of 2015~XX$_{169}$. As in the previous case, the object follows 
    an asymmetric horseshoe path with respect to the Earth with the value of the relative mean longitude librating around 180\degr, but 
    enclosing 0\degr (see Fig. \ref{shortya}, panel C). The short-term evolution of the value of its semi-major axis is rather irregular 
    (see Fig. \ref{shortya}, panel D) although it switched between the Apollo and Aten dynamical classes very recently and it will switch 
    back to Apollo in the future. The value of the Kozai parameter changes significantly over the time interval plotted (see Fig. 
    \ref{shortya}, panel B). The behaviour of eccentricity (see Fig. \ref{shortya}, panel E), inclination (see Fig. \ref{shortya}, panel F) 
    and argument of perihelion (see Fig. \ref{shortya}, panel G) is far from what is expected when a Kozai resonance is in operation. The 
    descending node is close to Earth's perihelion, not aphelion as in the case of 2015~XX$_{169}$, and the ascending node is away from the 
    path of any planet (see Fig. \ref{shortya}, panel H). However, this situation will be inverted in about 300 yrs from now.
%
%
    \begin{figure*}
      \centering
       \includegraphics[width=0.325\linewidth]{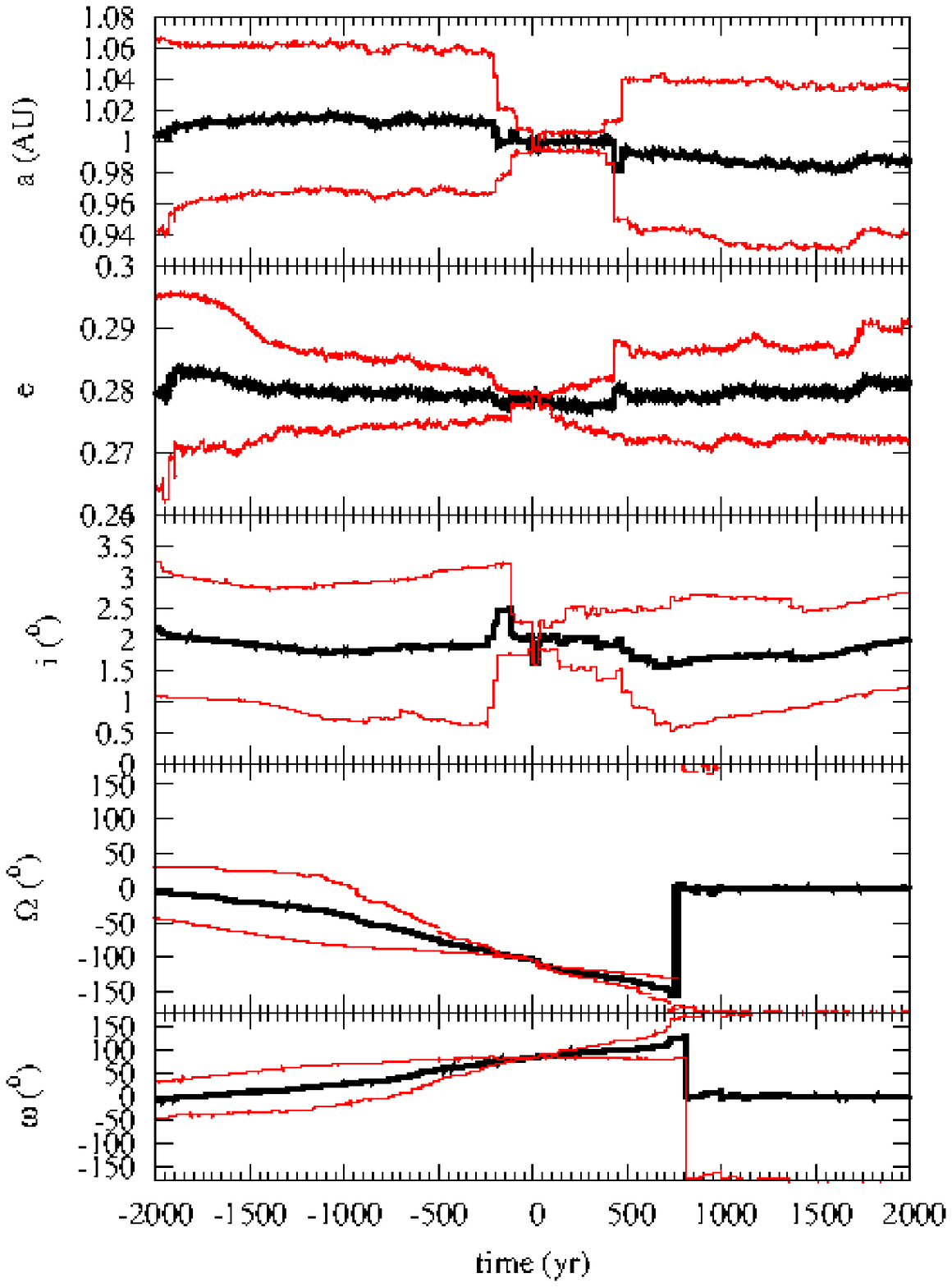}
       \includegraphics[width=0.325\linewidth]{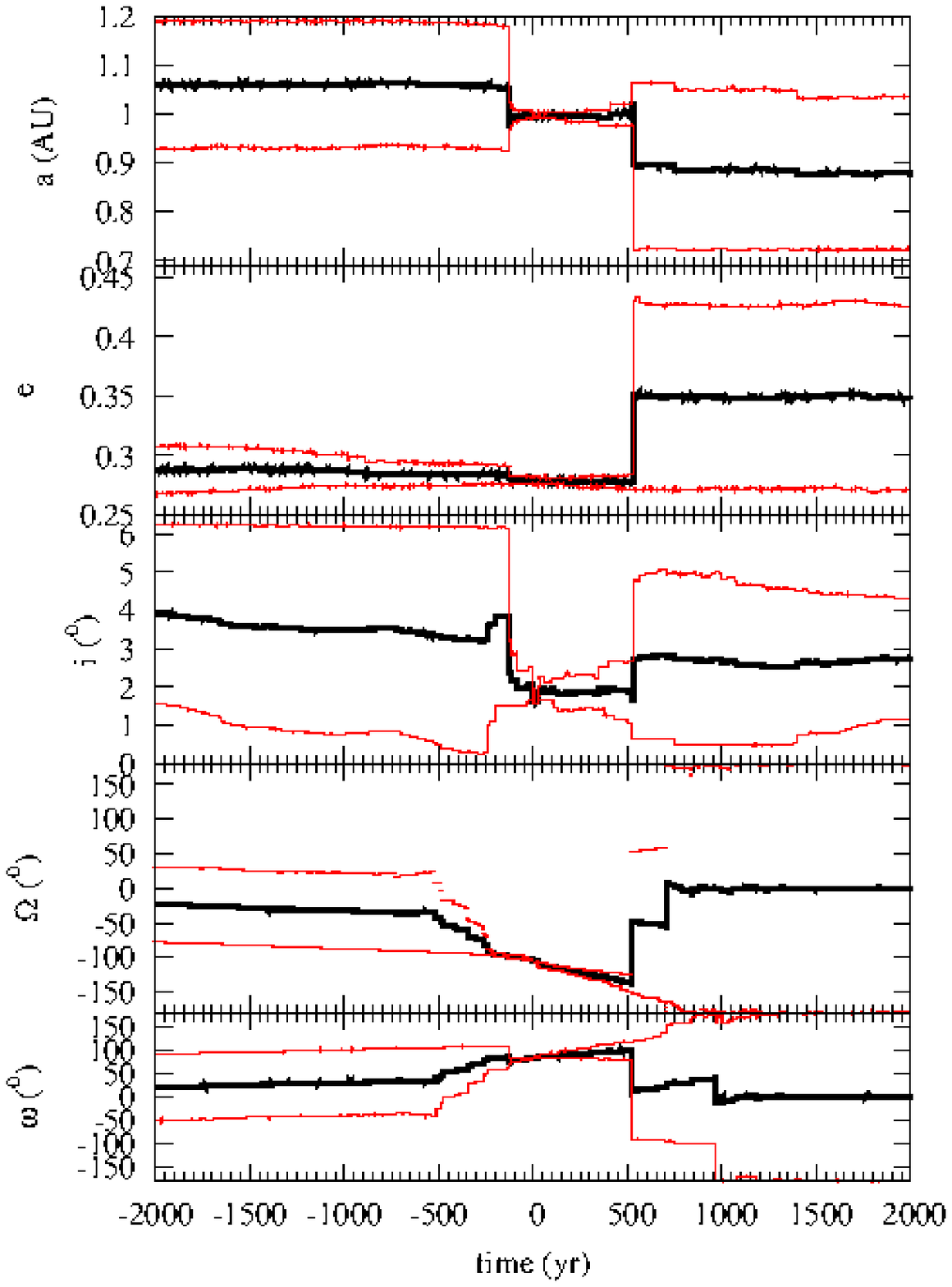}
       \includegraphics[width=0.325\linewidth]{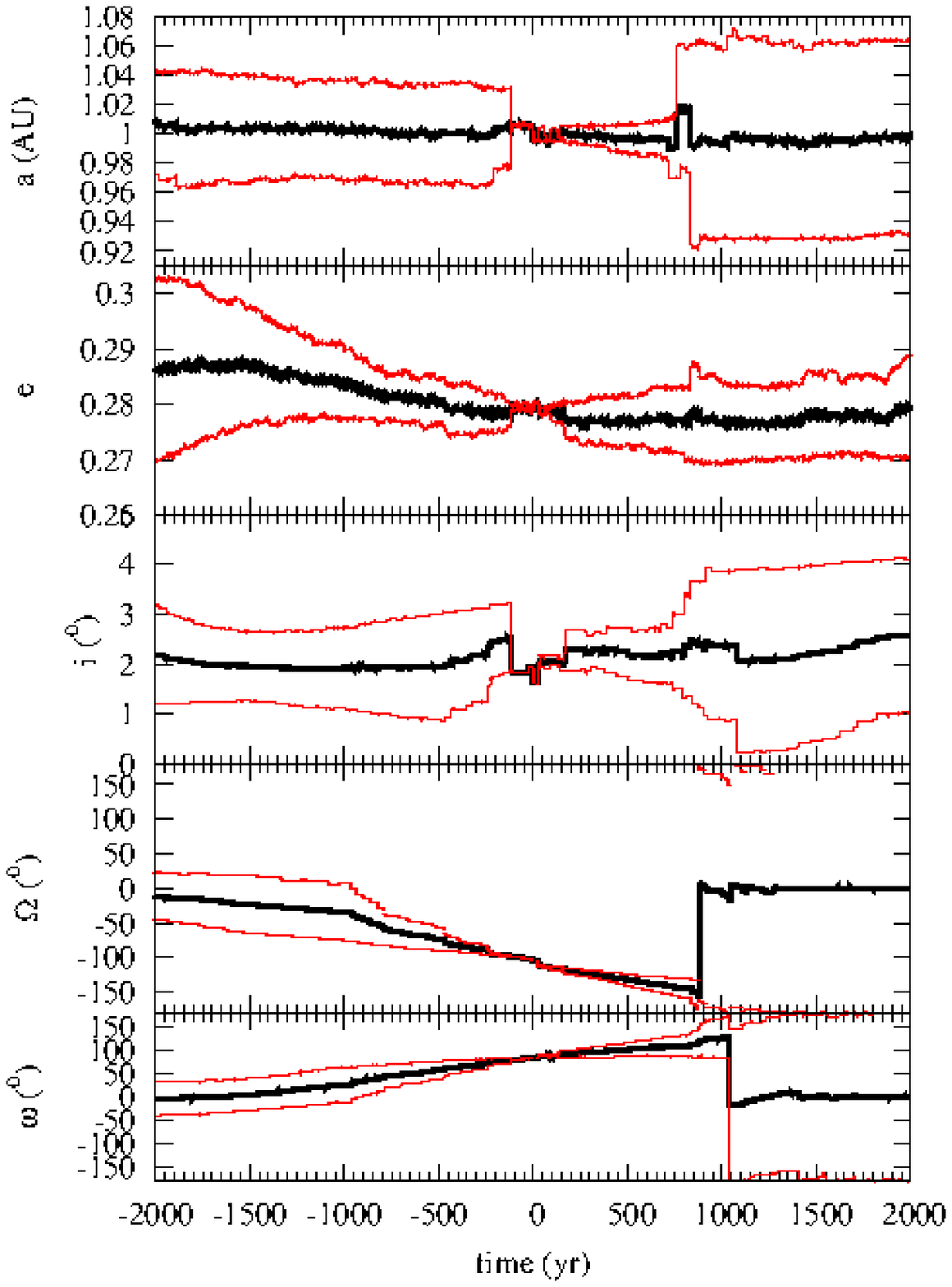}
       \caption{As Fig. \ref{errorsxx169} but for 2015~YA.
               }
       \label{errorsya}
    \end{figure*}
%
%

    The analysis of the effect of the errors in Table \ref{elements} on the evolution of the orbital parameters displayed in Fig. 
    \ref{errorsya} shows that the value of its Lyapunov time was very short ---below 100 yr--- in the past and it will be only slightly 
    longer in the future (a few hundred years). In this case, the average short-term orbital evolution predicted by the classical method 
    (left-hand panels in Fig. \ref{errorsya}) is consistent with that from the MCCM approach (right-hand panels in Fig. \ref{errorsya}).
%
%
     \begin{figure*}
       \centering
        \includegraphics[width=\linewidth]{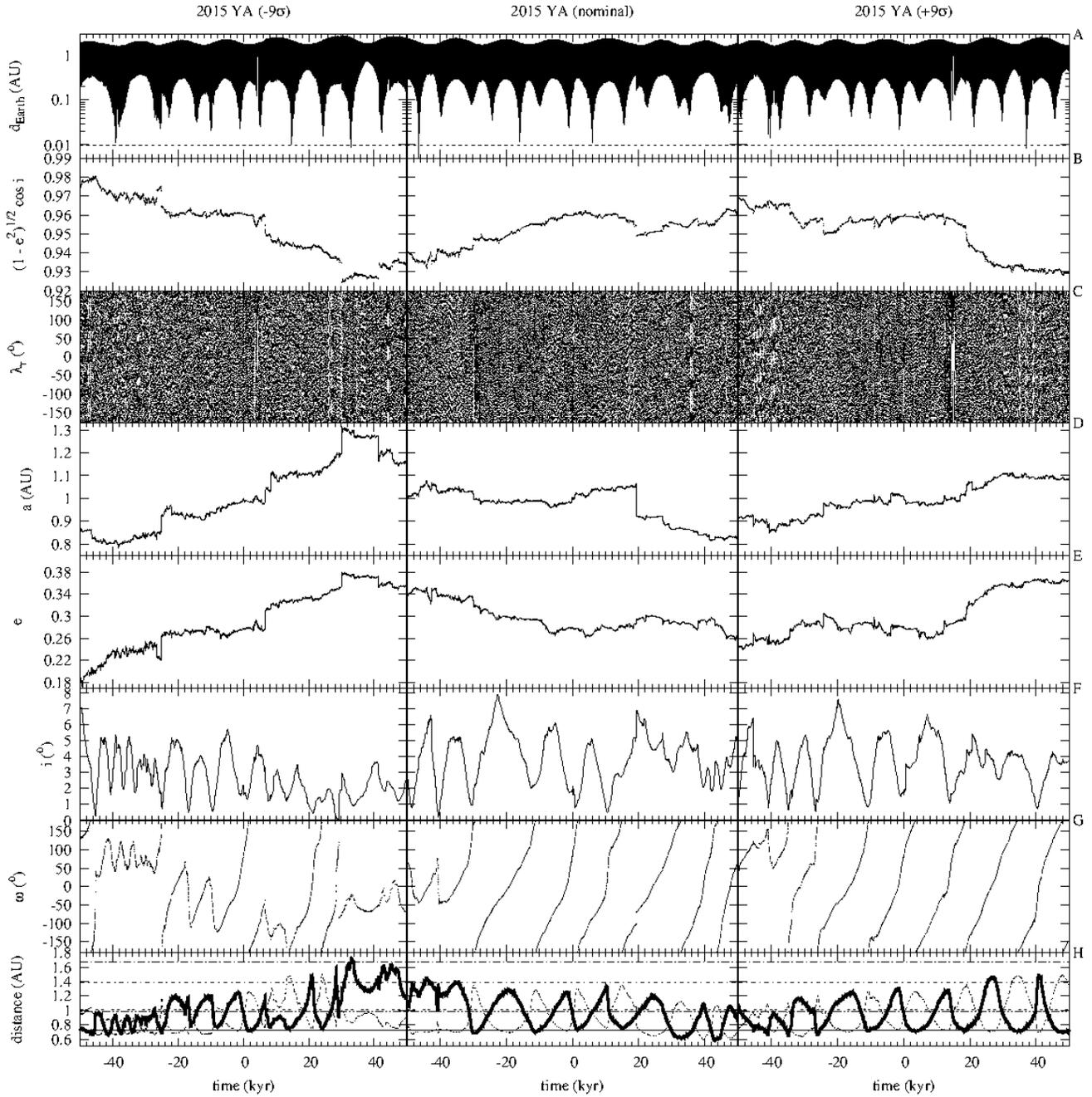}
        \caption{Same as Fig. \ref{longxx169} but for 2015~YA. Venus', Earth's and Mars' aphelion and perihelion distances are also shown in 
                 panel H. 
                }
        \label{longya}
     \end{figure*}
%
%

    The past long-term evolution of 2015~YA strongly suggests that its semi-major axis started to reach values similar to that of our planet 
    nearly 30 kyr ago (see Fig. \ref{longya}). It also includes phases in which the behaviour of the argument of perihelion is 
    consistent with what is expected when a Kozai resonance is in effect (see Fig. \ref{longya}, panel G) but the evolution of the 
    eccentricity (see Fig. \ref{longya}, panel E) does not exhibit the oscillatory behaviour that is observed, for instance, in the case of
    2015~XX$_{169}$. The positions of the nodes oscillate more or less regularly (see Fig. \ref{longya}, panel H) and the closest encounters
    with our planet often coincide with the time when both nodes are in the path of the Earth. The dynamics of this object is currently 
    controlled by encounters with Venus and the Earth-Moon system, but Mars may also be an important perturber in the future. 
%
%
     \begin{figure}
       \centering
        \includegraphics[width=\linewidth]{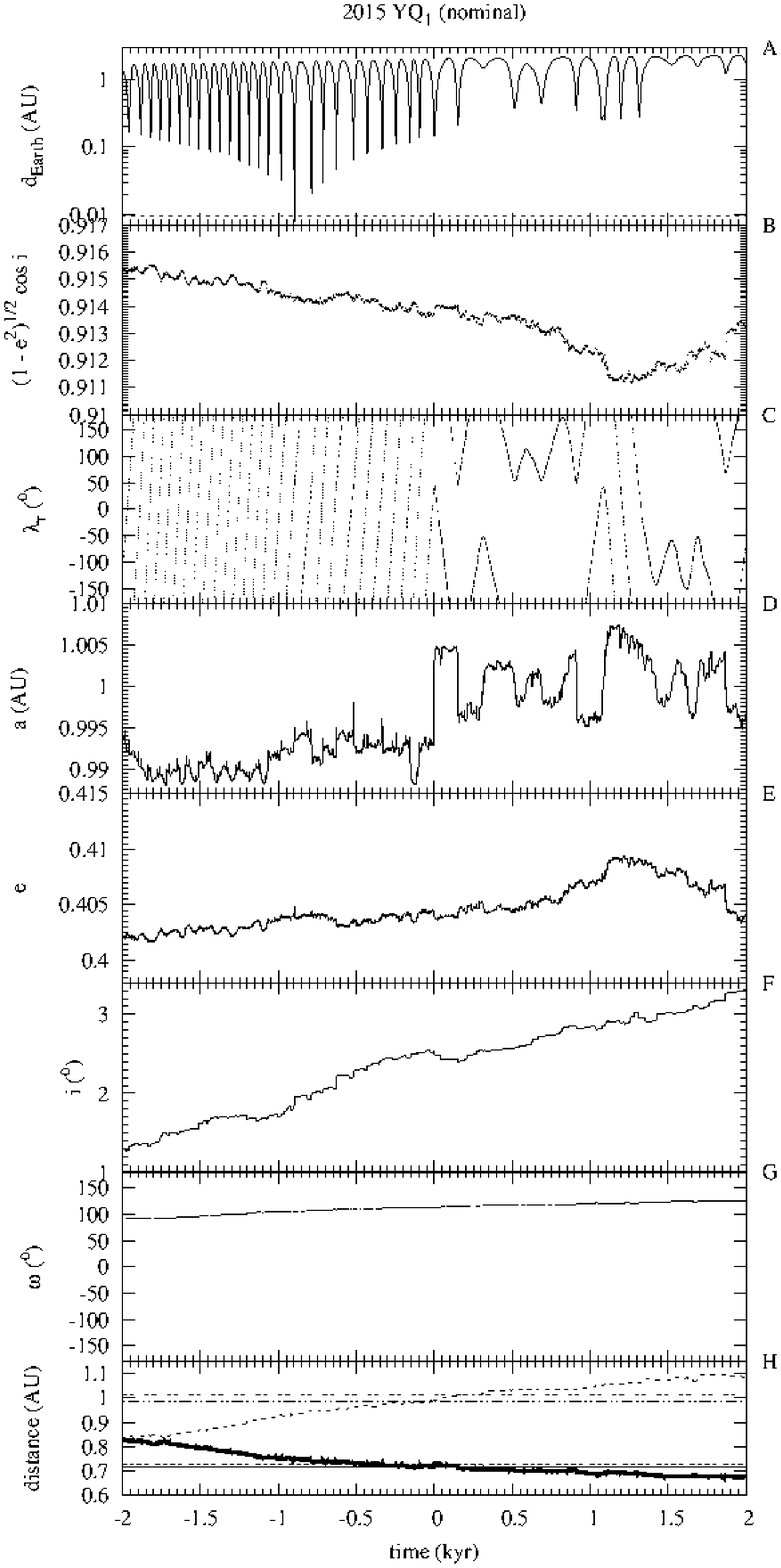}
        \caption{As Fig. \ref{shortxx169} but for 2015~YQ$_{1}$. 
                }
        \label{shortyq1}
     \end{figure}
%
%

 \section{Asteroid 2015~YQ$_{1}$: data and results}
    Asteroid 2015~YQ$_{1}$ was discovered on 2015 December 19 by A. D. Grauer observing for the Mt. Lemmon Survey (Bacci et al. 2015) at 
    $V=20.1$~mag. As the previous two objects, its orbital solution is in need of some improvement as it is currently based on 64 
    observations obtained during a time span of 3 d. It is the smallest of the trio of objects studied here with $H$ = 28.1 mag, which 
    translates into a diameter in the range 7--16 m for an assumed albedo of 0.20--0.04. Its orbit is also the most eccentric ($e=0.40398$) 
    of the set but its inclination is similar to that of 2015~YA ($i=2\fdg5$). With these values of the orbital parameters it may be the 
    most unstable of the trio. It currently belongs to the Apollo dynamical class. The value of its Earth MOID is 0.00052 AU and this NEA 
    experienced a close encounter with our planet on 2015 December 22 at 0.0037 AU. It undergoes close encounters with Venus, the Earth-Moon 
    system, and Mars at regular intervals. Like the previous two, it has been included in the NHATS list. Once more we have a dynamically 
    interesting small object with few observations; another obvious case of the situation described in the introduction section.  

    The immediate, past and future, orbital evolution of 2015~YQ$_{1}$ is displayed in Fig. \ref{shortyq1}; it follows an asymmetrical 
    horseshoe trajectory with respect to the Earth like the other two as $\lambda_{\rm r}$ librates around 180\degr but enclosing 0\degr 
    (see Fig. \ref{shortyq1}, panel C). The object switches between the Aten and Apollo dynamical classes but the evolution of its 
    semi-major axis is nearly as irregular as that of 2015~YA (see Fig. \ref{shortyq1}, panel D) although the transitions will last longer 
    than in the case of 2015~YA. The value of the Kozai parameter varies more than that of 2015~YA (compare Figs. \ref{shortya} and 
    \ref{shortyq1}, panel B) over the displayed time interval. Both eccentricity (see Fig. \ref{shortyq1}, panel E) and inclination (see 
    Fig.  \ref{shortyq1}, panel F) tend to increase over time, although the eccentricity decreases after 1200 yrs into the future. The value 
    of the argument of perihelion remains confined to the range (90, 175)\degr (see Fig. \ref{shortyq1}, panel G) which is somewhat 
    consistent with a Kozai resonance at high eccentricity, although not very strong. At present, the ascending node is close to Earth's 
    perihelion and the descending node is in the path of Venus (see Fig. \ref{shortyq1}, panel H).
%
%
    \begin{figure*}
      \centering
       \includegraphics[width=0.325\linewidth]{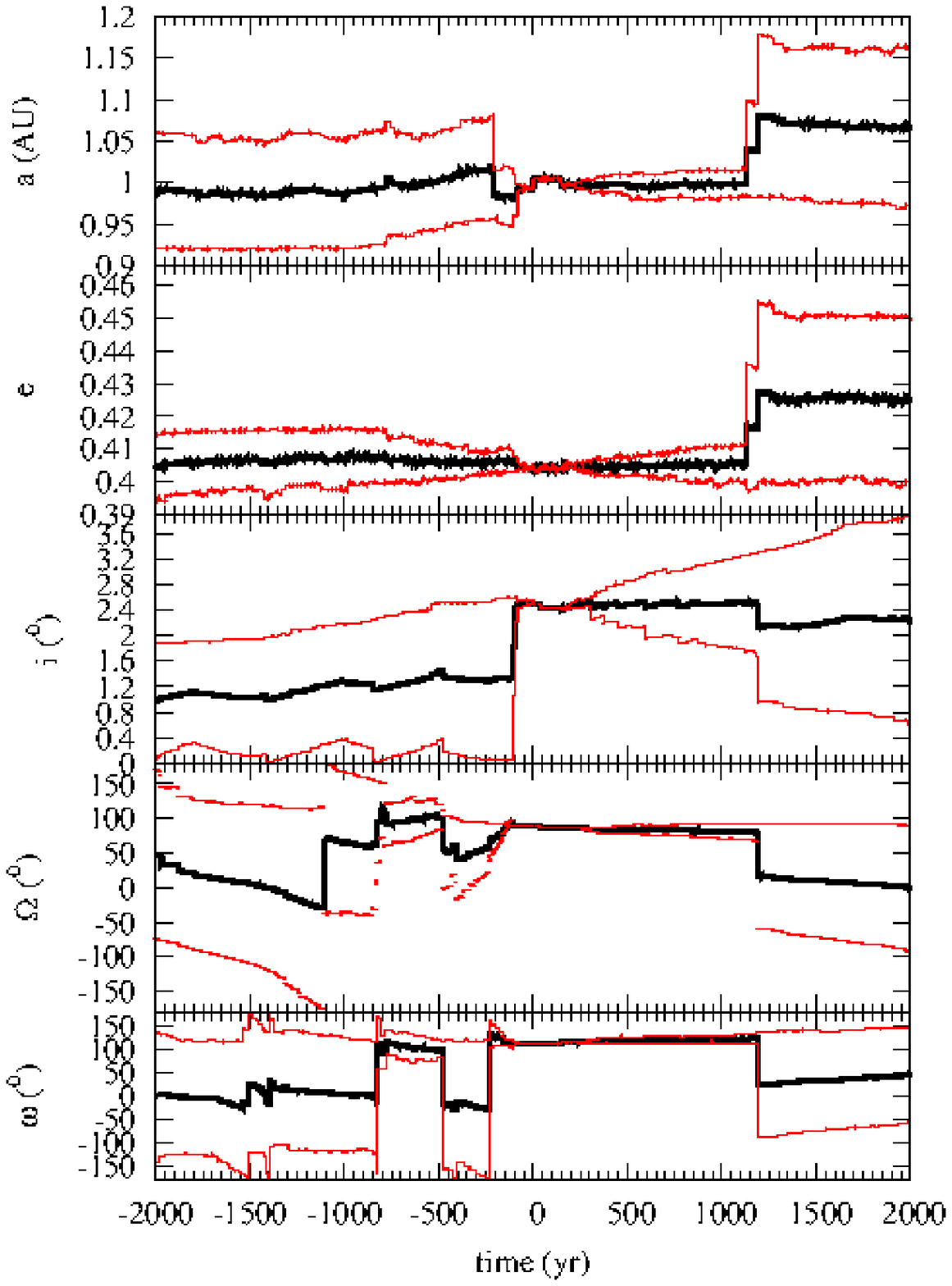}
       \includegraphics[width=0.325\linewidth]{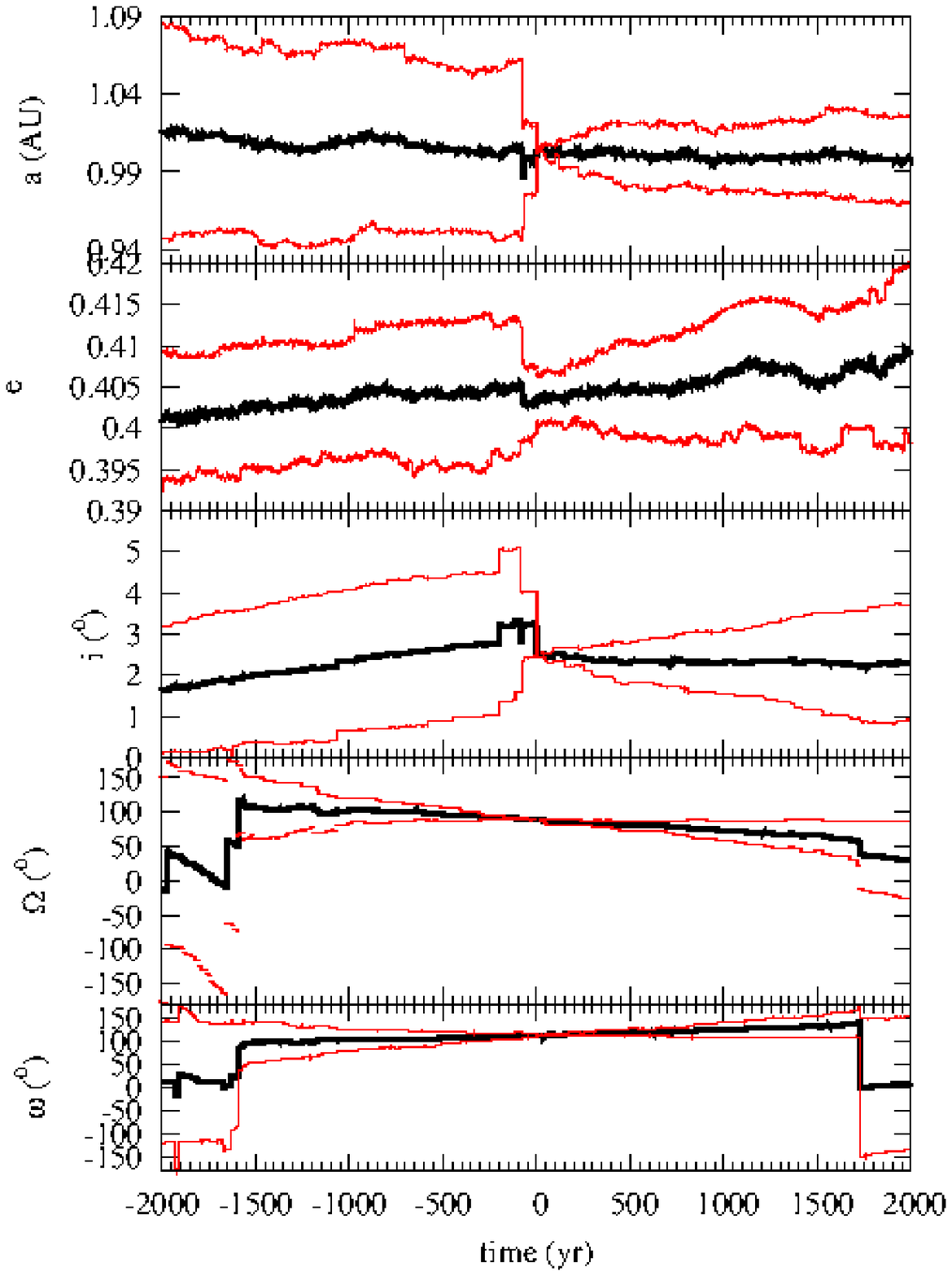}
       \includegraphics[width=0.325\linewidth]{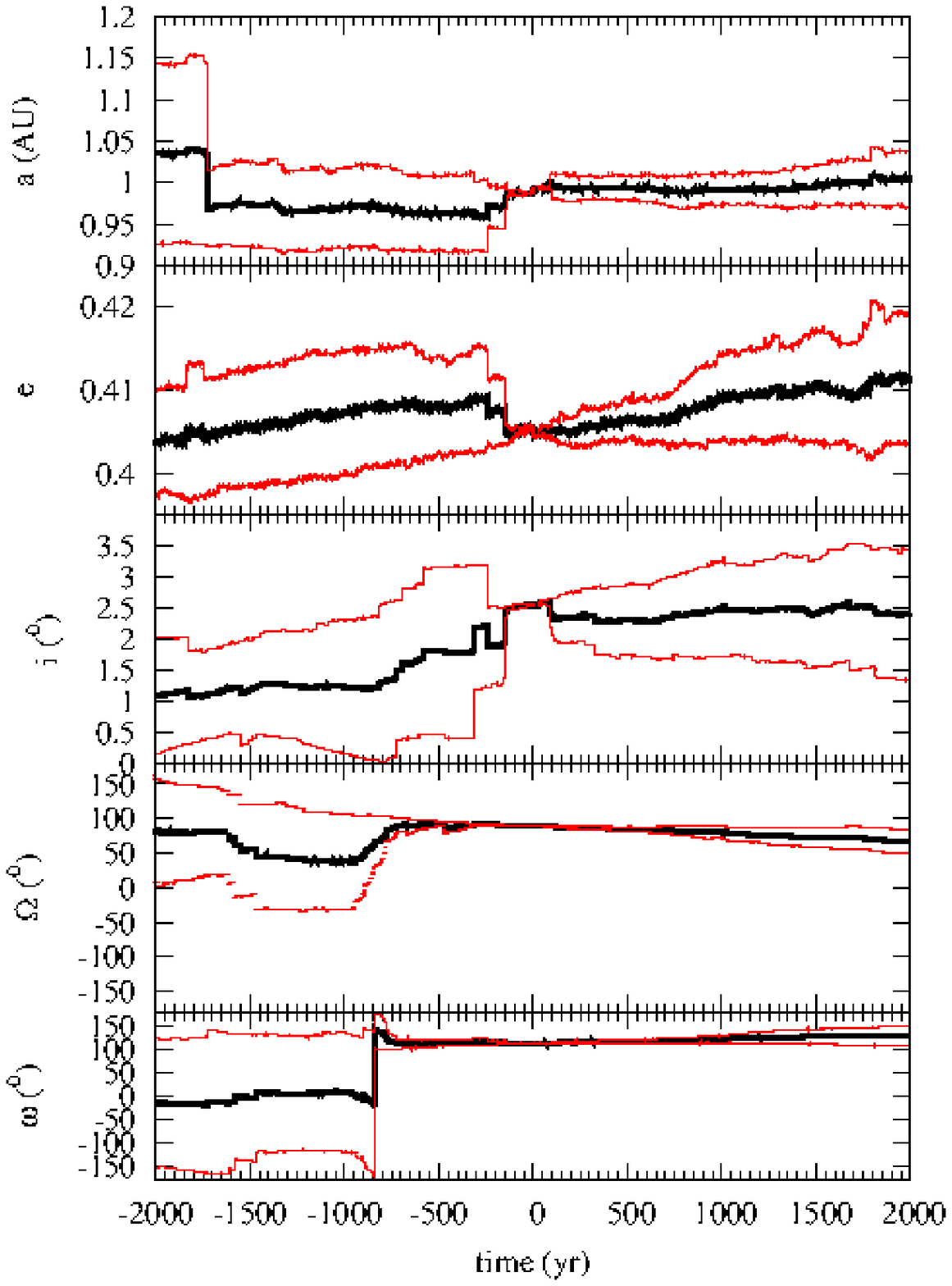}
       \caption{As Fig. \ref{errorsxx169} but for 2015~YQ$_{1}$. 
               }
       \label{errorsyq1}
    \end{figure*}
%
%
    Figure \ref{errorsyq1} shows that 2015~YQ$_{1}$ is probably the most dynamically unstable of the objects studied here. The data suggest 
    that the value of its Lyapunov time is very short, perhaps as low as a few decades, both for its past and future orbital evolution. The 
    average short-term orbital evolution predicted by the classical method (left-hand panels in Fig. \ref{errorsyq1}) is somewhat consistent 
    with that from the MCCM approach (right-hand panels in Fig. \ref{errorsyq1}) but its dynamical behaviour is difficult to predict beyond
    a century or so.
%
%
     \begin{figure*}
       \centering
        \includegraphics[width=\linewidth]{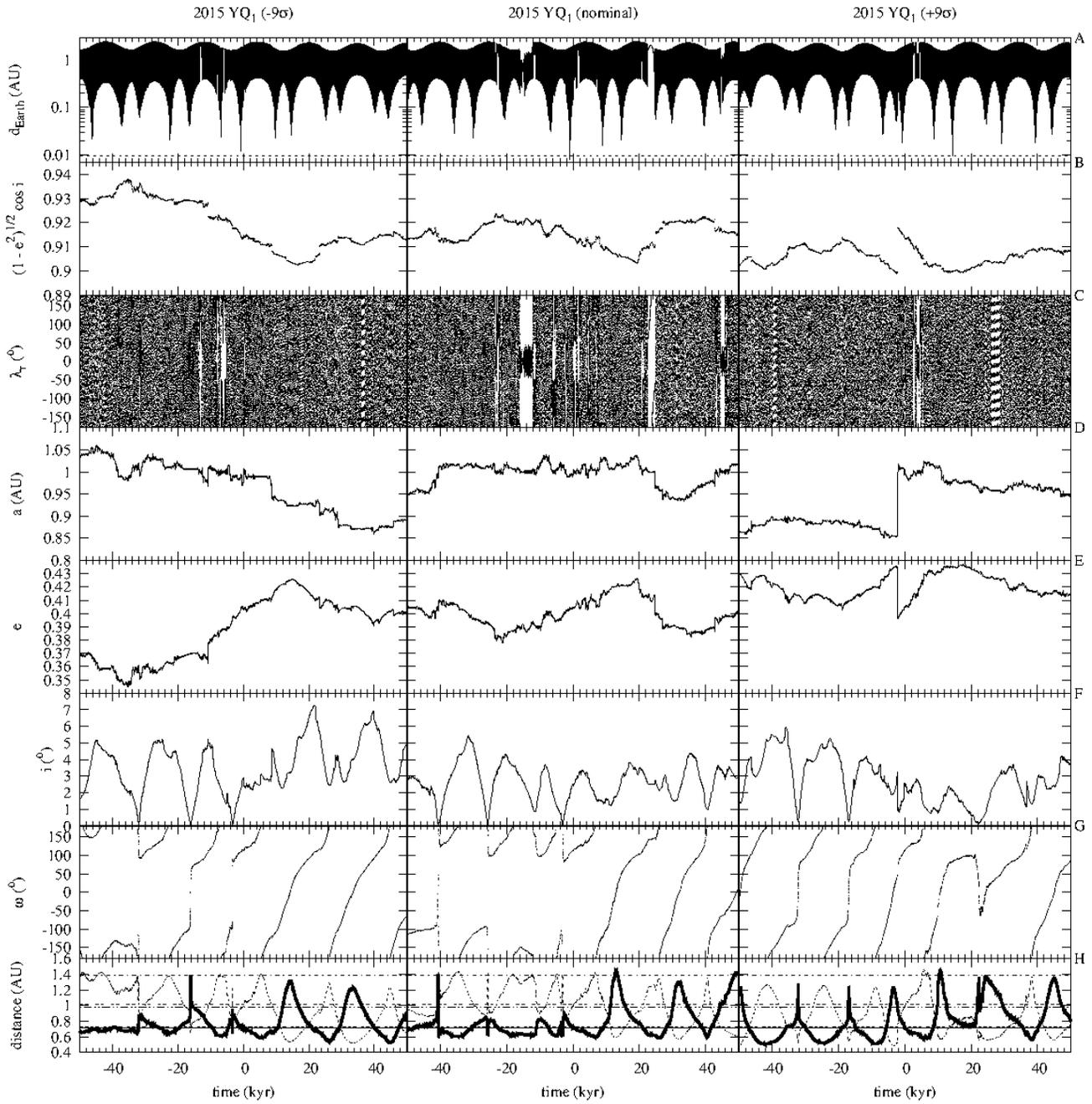}
        \caption{Same as Fig. \ref{longxx169} but for 2015~YQ$_{1}$. Venus', Earth's and Mars' aphelion and perihelion distances are also 
                 shown in panel H. 
                }
        \label{longyq1}
     \end{figure*}
%
%

    The past and future long-term dynamical evolution of 2015~YQ$_{1}$ shows (see Fig. \ref{longyq1}) that although it does not go beyond the 
    orbit of Mars, it travels relatively frequently inside that of Venus. As in the case of 2015~YA, it also experiences phases in which the 
    behaviour of the argument of perihelion is consistent with the one expected when a Kozai resonance is in effect (see Fig. \ref{longyq1}, 
    panel G) with libration about a value of 180\degr, for example. However, the evolution of both eccentricity and inclination (see Fig. 
    \ref{longyq1}, panels E and F) is rather irregular and does not exhibit the coupled oscillatory behaviour that is observed, for 
    instance, in the case of 2015~XX$_{169}$ (see Fig. \ref{longxx169}, panels E and F). The positions of the nodes oscillate more or less 
    regularly (see Fig. \ref{longyq1}, panel H) between 0.5 AU and 1.4 AU (Mars' perihelion). The dynamics of this object is controlled by 
    close encounters with Venus, the Earth-Moon system, and Mars.

 \section{Discussion}
    The three objects studied here were discovered within ten days, they are small objects with short data arcs, and move co-orbital to our
    planet. However, their orbits are quite diverse, mainly because of their eccentricities. As a result, their orbital evolution is also
    rather different. They have similar values of the semi-major axis and moderate to low inclinations. The Lyapunov time decreases as the
    eccentricity increases. 

    In general, Earth co-orbitals are small, for example 2003~YN$_{107}$, 2002~AA$_{29}$ or 2001~GO$_{2}$, but two of them are rather 
    large ---3753 Cruithne (1986 TO) has $H=15.7$~mag (Wiegert et al. 1997, 1998) and 85770 (1998~UP$_{1}$) has 
    $H=20.5$~mag.\footnote{\url{http://www.astro.uwo.ca/~wiegert/eca/}} Both are horseshoe librators like the NEAs discussed here. Other 
    NEAs experiencing horseshoe behaviour are 54509 YORP (2000 PH$_{5}$) (Wiegert et al. 2002; Margot \& Nicholson 2003), 2001~GO$_{2}$ 
    (Wiegert et al. 2002; Margot \& Nicholson 2003; Brasser et al. 2004), 2002~AA$_{29}$ (Connors et al. 2002; Brasser et al. 2004), 
    2003~YN$_{107}$ (Brasser et al. 2004; Connors et al. 2004), 2010~SO$_{16}$ (Christou \& Asher 2011), 2013~BS$_{45}$ (de la Fuente Marcos 
    \& de la Fuente Marcos 2013) and 2015~SO$_{2}$ (de la Fuente Marcos \& de la Fuente Marcos 2016). In contrast with what happens with 
    Jupiter where Trojans dominate, in the case of the Earth horseshoe librators outnumber any other co-orbital type with 12 objects 
    including the three studied here. Other Earth co-orbitals are 2010~TK$_{7}$ (Connors et al. 2011), a Trojan, and four quasi-satellites 
    ---164207 (2004~GU$_{9}$) (Connors et al. 2004; Mikkola et al. 2006; Wajer 2010), 277810 (2006~FV$_{35}$) (Wiegert et al. 2008; Wajer 
    2010), 2013~LX$_{28}$ (Connors 2014), and 2014~OL$_{339}$ (de la Fuente Marcos \& de la Fuente Marcos 2014). The existence of asteroids 
    moving in long-lasting horseshoe orbits associated with the Earth was first predicted by Hollabaugh and Everhart (1973). All the known 
    Earth co-orbitals are transient companions to our planet and most of them will remain as such for less than a few thousand years. Very 
    likely, the most stable of the currently known Earth co-orbitals is 2010~SO$_{16}$ that stays as horseshoe librator for at least 120 kyr 
    and possibly up to 1 Myr (Christou \& Asher 2011), remaining in the same co-orbital configuration during this time span. Asteroid 
    2015~SO$_{2}$ could be nearly as stable as 2010~SO$_{16}$, but it switches between co-orbital configurations on a regular basis (de la 
    Fuente Marcos \& de la Fuente Marcos 2016). Asteroid 2015~XX$_{169}$ is certainly less stable than 2015~SO$_{2}$, but more stable than 
    many of the objects listed above. Although co-orbitals are often considered as mere dynamical curiosities, some of them are relatively 
    easy to access from our planet (see, e.g., Stacey \& Connors 2009). Such objects are good candidates for in situ study, sample return 
    missions, or even commercial mining (e.g. Lewis 1996; Elvis 2012, 2014; Garc\'{\i}a Y\'arnoz et al. 2013; Harris \& Drube 2014). The 
    three objects studied here and several other Earth co-orbitals are included in the NHATS list of potentially accessible targets that 
    compiles attractive objects of interest for, e.g., future robotic missions.

    For NEOs with values of the semi-major axis close to that of our planet, mean-motion resonances other than the one with the Earth are 
    relatively weak, but secular resonances where the precession of the node of the perihelion of a NEO relative to a planet librates could 
    be strong. The effect of secular resonances for objects moving in orbits with semi-major axes smaller than 2~AU and relatively low 
    values of the eccentricity was first explored by Michel \& Froeschl\'e (1997) and further studied in Michel (1997, 1998). Michel \& 
    Froeschl\'e (1997) found that objects with $0.9<a<1.1$~AU are affected by the Kozai mechanism (Kozai 1962) that, at low inclination, 
    induces libration of the argument of perihelion around 0\degr or 180\degr (see, e.g., Michel \& Thomas 1996). An argument of perihelion 
    librating around 0\degr means that the orbit reaches perihelion at approximately the same time it crosses the Ecliptic from South to 
    North (the ascending node); a libration around 180\degr implies that the perihelion is close to the descending node. For higher 
    inclinations, the argument of perihelion usually oscillates around 90\degr or 270\degr (-90\degr) when the Kozai resonance is in effect. 
    Examples of both behaviours are found in our analyses above. The value of the argument of perihelion of 2015~XX$_{169}$ is found 
    librating about 270\degr when its inclination is high and about 180\degr when its inclination is low (see Fig. \ref{longxx169}, panel 
    G). For 2015~YA, librations about 0\degr, 90\degr or 270{\degr} are observed (see Fig. \ref{longya}, panel G) depending on the value of 
    the inclination. In the case of 2015~YQ$_{1}$ and due to its overall lower inclination, only libration around 0\degr or 180\degr is 
    observed (see Fig. \ref{longyq1}, panel G).

    Almost certainly, the three objects studied here are fragments of larger objects, which may also be fragments themselves. We can 
    speculate that this can be reconciled with a hypothetical origin in the Earth-Moon system as suggested by, e.g., Margot \& Nicholson 
    (2003) within the context of triggered resurfacing events when a larger asteroid encounters our planet at relatively large planetary 
    flyby distances, in the range 5--20 planetary radii (see, e.g., Keane \& Matsuyama 2015) or alteration of the spin rate (Scheeres et al. 
    2005) and subsequent rotationally induced structural failure (see, e.g., Denneau et al. 2015). Several hundred NEAs have values of the 
    MOID under 20 times the Earth radius. Within this hypothetical but plausible scenario, relatively large fragments can be released when  
    asteroids pass close to our planet. These fragments may remain within the orbital neighbourhood of our planet trapped in the web of 
    overlapping resonances that permeates the entire region (see, e.g., Christou 2000; de la Fuente Marcos \& de la Fuente Marcos 2015a) and 
    some of them may populate the co-orbital zone, in particular if the larger parent body was itself already a co-orbital. This scenario is 
    different from the original hypothesis of origin in the Earth-Moon system, which was by impact on the Moon (see, e.g., Warren 1994; 
    Gladman et al. 1995; Bottke et al. 1996; Gladman 1996).

    In our simulations, the role of the Yarkovsky and Yarkovsky--O'Keefe--Radzievskii--Paddack (YORP) effects (see, e.g., Bottke et al. 
    2006) has been ignored. The neglection of these effects has no impact on the evaluation of the present dynamical status of the NEAs 
    discussed here, but may affect both the reconstruction of their dynamical past and any predictions made regarding their future orbital
    evolution. Accurate modelling of the Yarkovsky force requires relatively precise knowledge of the physical properties of the objects 
    affected, which is obviously not the case here.

 \section{Conclusions}
    In this research, we have used $N$-body calculations and statistical analyses to study the orbital evolution of the recently discovered 
    NEAs 2015~XX$_{169}$, 2015~YA and 2015~YQ$_{1}$. The main conclusions of our study can be summarised as follows:
    \begin{itemize}
       \item Asteroids 2015~XX$_{169}$, 2015~YA and 2015~YQ$_{1}$ currently follow asymmetric horseshoe trajectories with respect to the 
             Earth (probability $>99.8$\%); they are small transient Earth co-orbitals.  
       \item Asteroid 2015~XX$_{169}$ is the most stable of the trio and it may remain in the vicinity of Earth's co-orbital region for many 
             thousands of years. Both 2015~YA and 2015~YQ$_{1}$ may not stay as Earth co-orbital companions for long.
       \item Asteroids 2015~XX$_{169}$, 2015~YA and 2015~YQ$_{1}$ are temporarily subjected to various forms of the Kozai resonance during
             the integrations performed. This secular resonance is stronger for 2015~XX$_{169}$.
       \item The known population of Earth co-orbitals is dominated by horseshoe librators, 12 out of 17 objects.
       \item Relatively poor orbital solutions can still be used to obtain robust conclusions on the dynamical evolution of NEOs if 
             extensive statistical sampling of the orbits is performed.
    \end{itemize}
    One of the objectives of this research is to bring these interesting NEAs to the attention of the astronomical community, encouraging 
    follow-up observations. Spectroscopic studies during their next perigee should be able to confirm if an origin in the Earth-Moon system 
    as suggested by, e.g., Warren (1994), Gladman et al. (1995), Bottke et al. (1996), Gladman (1996) or Margot \& Nicholson (2003) is 
    plausible.

 \acknowledgments
    We thank the referee, M. Connors, for his positive and helpful report, and S.~J.~Aarseth for providing the main code used in this 
    research. In preparation of this paper, we made use of the NASA Astrophysics Data System, the ASTRO-PH e-print server and the MPC data 
    server.

\end{document}